\documentclass[11pt]{article}
\def\article#1#2#3#4#5{
	\hrule				
	\vskip 2em
 	\begin{center}
  		{\LARGE \bf {#1} \par}		
  		\vskip 1.5em          	{\large {#2} \par}
  		\vskip 1em
  		{\large {#3 #4}				  
		\par
              {\tt  #5}}     		 
	\end{center}%
 	\par
	\hrule		     		
 	\vskip 1.5em
}

\newcounter{ewmsection}
\newcounter{ewmsubsection}[ewmsection]

\newcommand{\ewmsection}[1]{%
	\stepcounter{ewmsection}
	\section{\theewmsection . #1}
}
\newcommand{\ewmsubsection}[1]{%
	\stepcounter{ewmsubsection}
	\subsection{\theewmsection .\theewmsubsection . #1}
}

\usepackage{amsfonts}
\usepackage{amssymb}
\usepackage{amscd}
\newcommand{\C}{{\mathbb C}}

\newcommand{\ca}{{\mathcal A}}
\newcommand{\cb}{{\mathcal B}}
\newcommand{\ce}{{\mathcal E}}
\newcommand{\cf}{{\mathcal F}}
\newcommand{\cg}{{\mathcal G}}
\newcommand{\ch}{{\mathcal H}}

\newcommand{\cm}{{\mathcal M}}
\newcommand{\cp}{{\mathcal P}}
\newcommand{\cs}{{\mathcal S}}

\newcommand{\U}{{\mathcal U}}
\newcommand{\cz}{{\mathcal Z}}
\newcommand{\fb}{{\mathfrak B}}
\newcommand{\ff}{{\mathfrak f}}
\newcommand{\fri}{{\mathfrak i}}
\newcommand{\frg}{{\mathfrak G}}

\newcommand{\fg}{{\mathfrak g}}
\newcommand{\fs}{{\mathfrak S}}
\newcommand{\fu}{{\mathfrak U}}
\def\a{\alpha}
\def\b{\beta}
\def\g{\gamma}
\def\d{\delta}

\def\l{\lambda}
\def\p{\partial}
\def\th{\theta}

\def\vp{\varphi}

\def\z{\zeta}
\def\lra{\longrightarrow}
\def\hra{\hookrightarrow}
\renewcommand{\ker}{{\mathop{\mbox{Ker}}\nolimits\,}}

\begin{document}

\article{Moduli Space of Self-Dual Gauge Fields,\\
Holomorphic Bundles and Cohomology Sets\footnote{Invited talk
at the EWM Workshop on Moduli Spaces in Mathematics and Physics,
2-3 July 1998, Oxford, to appear in the Proceedings.}}
{Tatiana Ivanova}
{Laboratory of Theoretical Physics, JINR, Dubna,}
{Russia}
{ita@thsun1.jinr.ru}

{\bf Abstract.} We discuss the twistor correspondence between
complex vector bundles over a self-dual four-dimensional manifold
and holomorphic bundles over its twistor space and describe the
moduli space of self-dual Yang-Mills fields in terms of \v{C}ech
and Dolbeault cohomology sets. The cohomological description provides
the geometric interpretation of symmetries of the self-dual Yang-Mills
equations.

\ewmsection{Introduction}

The purpose of this paper is to describe the moduli space of
self-dual Yang-Mills fields and a symmetry algebra acting
on the solution space of the self-dual Yang-Mills equations.
The description of the moduli space of self-dual
Yang-Mills fields is based on the twistor construction~\cite{Pen,
W, AHS}.

Let us briefly outline the differential-geometric background.
We take $M$ to be an oriented Riemannian 4-manifold,
$G$  a semi-simple Lie group, $P(M,G)$ a principal fibre bundle
over $M$ with the structure group $G$, $A$ a connection 1-form on $P$,
$F_A$ its curvature 2-form and $D$ a covariant differential on $P$.
A connection 1-form $A$ on $P$ is called {\it self-dual} if its
curvature $F_A$ is {\it self-dual}, i.e.,
$$
*F_A=F_A,
\eqno(1)
$$
where $*$ is the Hodge star operator acting on 2-forms on $M$.
We shall call eqs.(1) the {\it self-dual Yang-Mills} (SDYM) equations.
By virtue of the Bianchi identity $D\,F_A=0$, solutions of the SDYM
equations automatically satisfy the Yang-Mills equations
$$
D(*F_A)=0.
\eqno(2)
$$

Notice that solutions to eqs. (2) are of considerable physical
importance (see the talk in this volume by S.T.Tsou~\cite{tsou}).
Physicists use
Yang-Mills theory (by which we mean any non-Abelian gauge theory) to
describe the strong and electroweak interactions (see
e.g.~\cite{CL}). They call the connection 1-form $A$ the {\it gauge
potential} and the curvature 2-form $F_A$ the {\it gauge} or {\it
Yang-Mills field}. The SDYM equations (1) describe a subclass of
solutions to the Yang-Mills equations (2). A choice of different
boundary conditions for self-dual gauge fields gives such important
solutions of the Yang-Mills equations as instantons, monopoles and
vortices.

It is well known that the SDYM equations are manifestly
invariant under the gauge transformations of the gauge potential
$A$ and gauge field $F_A$ and under the rescaling of a metric
{\bf g} on $M:$ {\bf g}$\mapsto e^\vp${\bf g} (Weyl transformation),
where $\vp$ is an arbitrary smooth function on $M$.
The gauge transformations have the form (cf.~\cite{tsou})
$$
A \mapsto A^g = g^{-1} A g + g^{-1} dg,
\eqno(3a)
$$
$$
F_A \mapsto F_A^g = g^{-1} F_A g,
\eqno(3b)
$$
where $g$ is a global section of the associated bundle of groups
Int$\,P=P\times_G G$ ($G$ acts on itself by internal automorphisms:
$h_1\mapsto h_2^{-1}h_1h_2,\ h_1,h_2\in G$), i.e., $g\in \Gamma (M,$
Int$\,P)$. We shall denote the infinite-dimensional Lie group
$\Gamma (M,$  Int$\,P)$ by $\frg_M$ and call it the {\it gauge group}.

Let us denote by $\ca_M$ the space of smooth global solutions
to eqs. (1). The {\it moduli space} $\cm$ of self-dual gauge
fields is the space of gauge nonequivalent self-dual gauge potentials
on $M$,
$$
\cm :=\ca_M/\frg_M.
\eqno(4)
$$
Let $U\subset M$ be such an open ball that the bundle $P$ is
trivializable over $U$. We consider smooth self-dual connection
1-forms $A$ on $U$, i.e., {\it local solutions} of the SDYM equations.
Denote by $\ca_U$ the space of all smooth solutions to eqs.(1)
on $U$ and by $\cm_U$ the {\it moduli space} of smooth self-dual
gauge potentials $A$ on $U$,
$$
\cm_U :=\ca_U/\frg_U,
\eqno(5)
$$
where $\frg_U:=\Gamma (U,$  Int$\,P)=C^\infty (U, G)$ is an
infinite-dimensional group of local gauge transformations.

The use of the moduli spaces (4) and (5) in physics is discussed
in the talk by S.T.Tsou~\cite{tsou}.
An important example of their use in mathematics is given by
Donaldson's discovery of exotic smooth structures on 4-manifolds,
which is based on topological properties of the moduli space of
self-dual gauge fields over the manifolds in question~\cite{D}.

The paper is organized as follows: in \S\,2 we recall the
twistor description of self-dual manifolds and self-dual
gauge fields, in \S\,3 we discuss the cohomological
description of the moduli space of self-dual gauge fields
mainly following the paper~\cite{adp}, and in \S\,4 we
describe the infinitesimal symmetries of the SDYM equations from
the cohomological point of view (see also~\cite{ita1, ita2}).

\ewmsection{An important tool: Twistors}

{\em Twistors were introduced by Penrose in order to translate the
massless free-field equations in space-time into
holomorphic structures on a related complex manifold known as a
twistor space.  The twistor theory is based on an integro-geometric
transformation which transforms complex-analytic data on the twistor
space to solutions of massless field equations.  Suggested
originally for the description of linear conformally invariant
equations, the twistor method has proved very fruitful for solving
nonlinear equations of general relativity and Yang-Mills theories.
Namely, the Penrose nonlinear graviton construction~\cite{Pen} gives
the general local solution of the self-dual conformal gravity
equations, and the Ward twistor interpretation of self-dual
gauge fields~\cite{W} gives the general local solution of the SDYM
equations on self-dual 4-manifolds $M$.}

\ewmsubsection{Twistor spaces}

{}For each oriented Riemannian 4-manifold $M$ one can
introduce the manifold
$$
\cz := P(M, SO(4))/U(2)\simeq P(M, SO(4))\times_{SO(4)}S^2,
$$
where $P(M, SO(4))$ is the principal $SO(4)$-bundle of oriented
orthogonal frames on $M$. So, the space $\cz$ is a bundle associated
to $P(M, SO(4))$ with the typical fibre ${\C}P^1\simeq S^2$ and the
canonical projection $\pi :\cz\to M$. The manifold $\cz$ is
called the {\it twistor space} of $M$.

A Riemannian metric {\bf g} is self-dual if the anti-self-dual part
of the Weyl tensor vanishes~\cite{Pen, AHS, W80}. Manifolds $M$
with self-dual metrics are called {\it self-dual}. In~\cite{Pen,
AHS} it was shown that the twistor space $\cz$ for such $M$ is  a
complex 3-manifold. In what follows, we shall consider a self-dual
manifold $M$ and the twistor space $\cz$ of $M$.

The Levi-Civita connection on $M$ generates the splitting of the
tangent bundle $T(\cz )$ into a direct sum
$$
T(\cz )=V\oplus H
\eqno(6)
$$
of the vertical $V=\ker\pi_*$ and horizontal $H$ distributions.
The complexified tangent bundle of $\cz$ can be splitted into
a direct sum
$$
T^{\C}(\cz )=V^{\C}\oplus H^{\C}=T^{1,0}\oplus T^{0,1}
\eqno(7)
$$
of subbundles of type (1,0) and (0,1). Analogously one can split
the complexified cotangent bundle  of $\cz$ into a direct sum of
subbundles $T_{1,0}$ and $T_{0,1}$. Using the standard complex
structure on $S^2\simeq{\C}P^1\hra\cz$, one obtains
$$
T^{\C}(\cz )=(V^{1,0}\oplus H^{1,0})\oplus(V^{0,1}\oplus H^{0,1}).
\eqno(8)
$$
The distribution $V^{0,1}$ is integrable.

Denote by $\{V_a\}$, $\{\bar V_a\}$, $\{\theta^a\}$  and
$\{\bar\theta^a\}$ ($a=1,2,3$) local frames for the bundles
$T^{1,0}$, $T^{0,1}$, $T_{1,0}$ and $T_{0,1}$, respectively.
Because of (8), each of the local frames is spanned by
horizontal (when $a=1,2$) and vertical (when $a=3$) parts.
The derivative operator $d$ on $\cz$ splits as follows:
$$
d=\p +\bar\p,\quad \p^2=\bar\p^2=0,\quad \p\bar\p +\bar\p\p=0,
\eqno(9)
$$
where locally $\p =\theta^aV_a$,  $\bar\p =\bar\theta^a\bar V_a$.

Let us consider a sufficiently small open ball $U\subset M$ such that
$\cz\mid_U$ is a direct product $\cp\equiv\cz\mid_U\simeq U\times S^2$
as a smooth real 6-manifold. The space $\cp\subset\cz$ is called the
{\it twistor space} of $U$. This space is covered by two coordinate
patches $\U_1$ and $\U_2$,
$$
\U_1:=U\times \Omega_1,\quad  \U_2:=U\times \Omega_2,
$$
where $\Omega_1=\{\l\in\C : |\l |<\infty\},\ \Omega_2=\{\z\in\C :
|\z |<\infty\}$  form the covering $\Omega =\{\Omega_1, \Omega_2\}$ of
the complex projective line $\C P^1$ and $\l =\z^{-1}$ on
$\Omega_1\cap\Omega_2$. On $\U_1$ and $\U_2$ we have the local
coordinates  $\{x^{\mu}, \l , \bar\l\}$ and  $\{x^{\mu}, \z , \bar\z\}$,
respectively. We denote by $\fu =\{\U_1, \U_2\}$ the two-set open
covering of $\cp =\U_1\cup\U_2$ and by $\U_{12}$ the intersection
$\U_1\cap\U_2=U\times (\Omega_1\cap\Omega_2)$.

Recall that for any self-dual manifold its twistor space is a complex
manifold. So, on $\U_1, \U_2\subset\cp$ one can introduce holomorphic
coordinates $\{z^a_1\},\ \{z^a_2\},\ a=1,2,3.$ On the intersection
$\U_{12}=\U_1\cap\U_2$ these coordinates are related by a holomorphic
transition function $f_{12}: z^a_1=f^a_{12}(z^b_2)$. For local frames
$\{\bar V^{(1)}_a\}$  and  $\{\bar V^{(2)}_a\}$ of the bundle $T^{0,1}$
over $\U_1$ and $\U_2$  one has $\bar V^{(1)}_a z^b_1 =0$
on $\U_1$ and  $\bar V^{(2)}_a z^b_2 =0$ on $\U_2$. Notice that as
local frames of  $T^{0,1}$ over $\U_1$, $\U_2$ one can take the
antiholomorphic vector fields $\{\frac{\p}{\p\bar z^a_1}\}$ on $\U_1$
and $\{\frac{\p}{\p\bar z^a_2}\}$ on $\U_2$.

\ewmsubsection{Twistor correspondence}

Let $M$ be a self-dual 4-manifold with the twistor space $\cz$.
There is a bijective correspondence~\cite{W, AW, AHS} between
complex vector bundles $E\to M$  on $M$ with self-dual connections
and holomorphic vector bundles $\tilde E\to\cz$ on $\cz$ which are trivial
on fibres $\C P^1$ of the bundle $\pi : \cz\to M$ (see
also~\cite{PenR, WW, MW} and references therein).

Let us briefly describe the twistor correspondence for the
case of a vector bundle $\ce$ over an open set $U\subset M$
with a self-dual connection 1-form $A$. Such a bundle
$(\ce , A)$ can be lifted to a bundle $(\pi^*\ce , \pi^*A)$
over the twistor space $\cp$ of $U$. By definition of the
pull-back, the pulled back connection 1-form $\pi^*A$ on
$\pi^*\ce$ is flat along the fibres $\C P^1$ of the bundle
$\pi : \cp\to U$.  Therefore, the components of the
connection 1-form $\pi^*A$ on the bundle $\tilde\ce_0:= \pi^*\ce$
along the distribution $V$ can be set equal to zero.
Moreover, the bundle $\tilde\ce_0$ is a trivial complex vector bundle
$\tilde\ce_0=\cp\times \C^n$ with the transition matrix
$\cf^0_{12}=1$ on $\U_1\cap\U_2$. As it was demonstrated in~\cite{W,
AW, AHS}, the SDYM equations (1) on a connection 1-form $A$ on $\ce$
is the condition for the connection 1-form $\pi^*A$ to define
a {\it holomorphic structure} on the bundle $\tilde\ce_0$. Namely,
the 1-form $\pi^*A$ can be splitted into a direct sum of $(1,0)$-
and $(0,1)$-parts, and the operator $\bar\p$ can be lifted from $\cp$
to $\tilde\ce_0$,
$$
\bar\p_{\bar B}=   \bar\p + \bar B,
\eqno(10)
$$
where $\bar B$ is the $(0,1)$-part of  $\pi^*A$  satisfying the equations
$$
\bar\p^2_{\bar B}\equiv   \bar\p\bar B + \bar B\wedge \bar B=0.
\eqno(11)
$$
In the local frame $\{\bar\th^a\}$, $a=1,2,3$, we have
$\bar B=\bar B_a\bar\th^a$ and $\bar B_3=0$. Let us denote the described
correspondence as $(\ce , A)\sim (\tilde\ce_0, \bar B)$. From (11)
it follows that the trivial holomorphic vector bundle $\tilde\ce_0$ with
the flat $(0,1)$-connection $\bar B$ is diffeomorphic to a
{\it holomorphic vector bundle} $\tilde\ce$ with a holomorphic
transition matrix  $\cf_{12}$, i.e.,  $(\tilde\ce_0, \bar B)\sim (\tilde
\ce ,\cf_{12})$.  Therefore, there exist smooth $G$-valued functions
$\psi_1$ on $\U_1$ and $\psi_2$ on $\U_2$ such that
$\bar B^{(1)}_a= -(\bar V^{(1)}_a\psi_1)\psi_1^{-1}$,
$\bar B^{(2)}_a= -(\bar V^{(2)}_a\psi_2)\psi_2^{-1}$ and
$\cf_{12}=\psi_1^{-1}\cf^0_{12}\psi_2= \psi_1^{-1} \psi_2$,
where $\cf^0_{12}=1$ is the transition matrix in the bundle
$\tilde\ce_0$. Since $\bar B$ is zero along the distribution
$V^{0,1}$, we have $\bar V^{(1)}_3\psi_1=0$
on $\U_1$ and  $\bar V^{(2)}_3\psi_2=0$ on $\U_2$, which means that
$\tilde\ce$ is holomorphically  trivial after the restriction to any
projective line $\C P^1_x\hra\cp$, $x\in U$.

To sum up, we have a one-to-one correspondence between the complex
vector bundle $\ce$ over $U\subset M$ with a self-dual connection
1-form $A$ and the trivial complex vector bundle $\tilde\ce_0$ over
$\cp$ with the flat $(0,1)$-connection $\bar B$ on $\tilde\ce_0$
having zero component along the distribution $V^{0,1}$.
In its turn, there is a diffeomorphism between the bundle $(\tilde\ce_0,
\bar B)$ and the holomorphic vector bundle $\tilde\ce$ over $\cp$ that
is trivializable as a smooth bundle over $\cp$ and is holomorphically
trivializable after restricting to $\C P^1_x\hra\cp$, $x\in U$.
Thus we have the following equivalence of data:
$$
(\ce , A)\sim (\tilde\ce_0, \bar B)\sim (\tilde\ce , \cf_{12}),
$$
which is called the twistor correspondence between the bundles
$(\ce , A)$, $(\tilde\ce_0, \bar B)$ and $(\tilde\ce , \cf_{12})$.


\newpage

\ewmsection{\v{C}ech and Dolbeault descriptions of holomorphic
bundles}

{\em In the \v{C}ech approach holomorphic bundles are described by
holomorphic transition matrices, and in the Dolbeault approach they
are described by flat (0,1)-connections.  In this section we recall
definitions of cohomology sets of manifolds with values in sheaves of
groups and reformulate the equivalence of the \v{C}ech and Dolbeault
descriptions of holomorphic bundles in cohomology terms. At last,
using the twistor correpondence, we obtain two cohomological
descriptions of the moduli space $\cm_U$ of self-dual gauge fields.}

\ewmsubsection{Sheaves and cohomology sets}

Let us recall some definitions~\cite{GR, Oni}.  We consider a complex 
manifold $X$, smooth maps from $X$ into a non-Abelian group $G$ and a 
sheaf $\fs$ of such $G$-valued functions.  Let $\fu =\{\U_\a\}, \a\in 
I$, be an open covering of the manifold $X$. A {\it q-cochain} of the 
covering $\fu$ with values in $\fs$ is a collection $\psi 
=\{\psi_{\a_0...\a_q}\}$ of sections of the sheaf $\fs$ over nonempty 
intersections $\U_{\a_0}\cap\dots\cap\U_{\a_q}$.  A set of q-cochains 
is denoted by $C^q(\fu ,\fs )$; it is a group under the pointwise 
multiplication.

Subsets of {\it cocycles} $Z^q(\fu ,\fs )\subset C^q(\fu ,\fs )$ for
$q=0,1$ are defined as follows
$$
Z^0(\fu,\fs):=\{\psi \in C^0(\fu,\fs):
\psi _\a \psi _\b^{-1}=1\ \mbox{on}\
\U_\a\cap\U_\b\ne\varnothing\},
\eqno(12a)
$$
$$
Z^1(\fu,\fs):=\left\{\psi \in C^1(\fu,\fs): \psi _{\b\a}=
\psi _{\a\b}^{-1}\
\mbox{on}\
\U_\a\cap\U_\b\ne \varnothing ;\right .
$$
$$
{\hspace{2cm}}\left .\psi _{\a\b}\psi _{\b\g}\psi _{\g\a}=1\ \mbox{on}\
\U_\a\cap\U_\b\cap\U_\g\ne\varnothing\right\}.
\eqno(12b)
$$
It follows from (12a) that $Z^0(\fu,\fs)$ coincides with the group
$H^0(X,\fs ):=\fs (X)\equiv\Gamma (X,\fs )$ of global sections of
the sheaf $\fs$. The set $Z^1(\fu,\fs)$ is not in general a subgroup
of the group $C^1(\fu,\fs)$.

Cocycles $\hat f, f\in Z^1(\fu,\fs)$ are called {\it equivalent}
$\hat f\sim f$ if $\hat f_{\a\b}=\psi_\a f_{\a\b}\psi _\b^{-1}$ for
some $\psi\in C^0 (\fu,\fs)$, $\a ,\b\in I$.
The cocycle $f$  equivalent to $\hat f=1$ is called {\it trivial}
and for such cocycles $f=\{f_{\a\b}\}$ we have $f_{\a\b}=
\psi_{\a}^{-1}\psi_{\b} $. A set of equivalence classes of
1-cocycles is called the {\it 1-cohomology set} and denoted by
$H^1(\fu ,\fs )$.   After taking the direct limit of the sets
$H^1(\fu ,\fs )$ over successive refinement of the covering $\fu$ of $X$,
one obtains the {\it \v{C}ech 1-cohomology set} $H^1(X,\fs )$ of $X$
with coefficients in $\fs$. In the case when $\fu_\a$ are Stein
manifolds, $H^1(\fu ,\fs )=H^1(X,\fs )$.

We shall also consider a  sheaf $\dot\fs$ of smooth functions on $X$
with values in an Abelian group. Then the subgroups of cocycles
$Z^q(\fu ,\dot\fs )\subset C^q(\fu ,\dot\fs )$ for $q=0,1$ are defined
as follows
$$
Z^0(\fu,\dot\fs):=\{\th \in C^0(\fu,\dot\fs): \th _\a -\th _\b=0\
\mbox{on}\
\U_\a\cap\U_\b\ne\varnothing\},
\eqno(13a)
$$
$$
Z^1(\fu,\dot\fs):=\left\{\th \in C^1(\fu,\dot\fs):
\th _{\a\b}+\th _{\b\a}=0\
\mbox{on}\
\U_\a\cap\U_\b\ne \varnothing ;\right .
$$
$$
{\hspace{2cm}}\left .\th _{\a\b}+\th _{\b\g}+\th _{\g\a}=0\ \mbox{on}\
\U_\a\cap\U_\b\cap\U_\g\ne\varnothing\right\},
\eqno(13b)
$$
i.e., everywhere in definitions the multiplication is replaced by addition.
Trivial cocycles (coboundaries) are given by the formula $\th_{\a\b}=
\th_\a -\th_\b$, where $\{\th_{\a\b}\}\in Z^1(\fu ,\dot\fs)$,
$\{\th_{\a}\}\in C^0(\fu ,\dot\fs)$.  Quotient spaces
(cocycles/coboundaries) are the cohomology spaces $H^i(\fu,\dot\fs ),
i=1,2,\dots\ $.

Now we consider the twistor space $\cp$ and the two-set open covering
$\fu=\{\U_1,\U_2\}$ of $\cp$. Then the space of cocycles $Z^1(\fu,\fs)$
with coefficients in a sheaf $\fs$ of non-Abelian groups over $\cp$
is a special case of formula (12b),
$$
Z^1(\fu,\fs):=\{f\in C^1(\fu,\fs):
f_{21}=f^{-1}_{12}\ \mbox{on}\ \U_1\cap\U_2\}.
\eqno(14)
$$
Any cocycle $f=\{f_{12}, f_{21}\}\in Z^1(\fu,\fs)$ defines
a unique complex vector bundle $\tilde\ce$ over $\cp=\U_1\cup\U_2$
by glueing the direct products $\U_1\times\C^n$ and $\U_2\times\C^n$
with the help of $G$-valued transition matrix $f_{12}$ on $\U_{12}$.
Equivalent cocycles define isomorphic complex vector bundles over $\cp$
and smooth complex vector bundles are parametrized by the
set $H^1(\cp,\fs)$.

Let us introduce the sheaf $\ch$ of all holomorphic sections
of the trivial bundle $\cp\times G$, where $G$ is a Lie group.
Then holomorphic vector bundles over $\cp$ are parametrized
by the set $H^1(\cp,\ch)$.

\ewmsubsection{Exact sequences of sheaves and cohomology sets}

We consider the sheaf $\fs$ of smooth sections of the bundle
$\cp\times G$  and the subsheaf $\cs\subset\fs$ of such smooth sections
that are annihilated by the distribution $V^{0,1}$ on $\cp$,
i.e., locally $\bar V_3\psi =0$ on $\U\subset\cp$. So we have
$\ch\subset\cs\subset\fs$ and there is the canonical embedding
$\fri: \ch\to\cs$.

Let us also consider the sheaf $\cb^{0,1}$ of such smooth
$(0,1)$-forms $\bar B$ on $\cp$ with values in the Lie algebra
$\cg$ of $G$ that have zero components along the distribution $V^{0,1}$.
Let us define a map $\bar\d^0: \cs\to\cb^{0,1}$ given for any open set
$\U\subset\cp$ by the formula
$$
\bar\d^0\psi = - (\bar\p\psi )\psi^{-1},
\eqno(15)
$$
where $\psi\in \cs(\U)$, $\bar\d^0\psi\in \cb^{0,1}(\U)$, $d=\p +\bar\p$.
One can also consider the sheaf  $\fb^{0,2}$ of smooth $\cg$-valued
$(0,2)$-forms on $\cp$ and introduce an operator $\bar\d^1: \cb^{0,1}\to
\fb^{0,2}$ defined for any open set $\U\subset\cp$ by the formula
$$
\bar\d^1 \bar B=\bar\p \bar B+\bar B\wedge\bar B,
\eqno(16)
$$
where $\bar B\in \cb^{0,1}(\U)$,  $\bar\d^1 \bar B\in\fb^{0,2}(\U)$.

Denote by $\cb$ the subsheaf in $\cb^{0,1}$ of such $\bar B$ that
$\bar\p \bar B+\bar B\wedge\bar B=0$, i.e., $\cb =\ker\bar\d^1$.
The sheaf $\cs$ acts on the sheaf $\cb$ by means of the adjoint
representation:
$$
\bar B\mapsto \mbox{Ad}_\psi \bar B=
\psi^{-1}\bar B\psi + \psi^{-1}\bar\p\psi .
$$
It can be checked that the sequence of sheaves
$$
{\bf 1}\lra\ch\stackrel{\fri}{\lra}\cs\stackrel{\bar\d^0}{\lra}
\cb\stackrel{\bar\d^1}{\lra}0
\eqno(17)
$$
is exact, i.e., $\cb\simeq\cs/\ch$.  The exact sequence of sheaves
induces the following exact sequence of cohomology sets~\cite{Oni, adp}:
$$
e\lra H^0(\cp, \ch)\stackrel{\fri_*}{\lra}H^0(\cp, \cs)
\stackrel{\bar\d^0_*}{\lra}H^0(\cp,\cb)\stackrel{\bar\d^1_*}{\lra}
H^1(\cp,\ch)\stackrel{\ff}{\lra} H^1(\cp, \cs) ,
\eqno(18)
$$
where $e$ is a marked element of these sets and $\ff$ is an embedding
induced by the map $\fri$.

The sets $H^0(\cp,\ch)$,  $H^0(\cp,\cs)$ and $H^0(\cp,\cb)$ are
the spaces of global sections of the sheaves $\ch, \cs$ and $\cb$.
The set $H^1(\cp,\ch)$ is the moduli space of holomorphic vector
bundles over $\cp$, and the set $H^1(\cp,\cs)$ is the moduli space of
smooth complex vector bundles over $\cp$ that are holomorphic on any
projective line $\C P^1_x\hra\cp$, $x\in U$.

\ewmsubsection{Cohomological description of the moduli space $\cm_U$}

By definition the moduli space $\cm_U$ of local solutions to the
SDYM equations is the space of gauge nonequivalent self-dual
connections $A$ on $U$ (see (5)). The space $H^0(\cp,\cb)$ is the
space of smooth $\cg$-valued global $(0,1)$-forms $\bar B$ on $\cp$
satisfying (11) and having zero component along the distribution
$V^{0,1}$. By virtue of the twistor correspondence $(\ce, A)\sim
(\tilde\ce_0, \bar B)$, the space $H^0(\cp,\cb)$ coincides with
the space $\ca_U$ of local solutions to the SDYM equations,
$H^0(\cp,\cb)\simeq\ca_U$. The group $H^0(\cp,\cs)$ is isomorphic
to the group $\frg_U$ of local gauge transformations, because
$G$-valued smooth functions $\psi$ defined globally on
$\cp =U\times\C P^1$ and holomorphic on $\C P^1$ do not depend on
local complex coordinates of $\C P^1$, i.e., $\psi\equiv g(x)\in\frg_U$,
$x\in U$. Therefore we have the bijection
$$
\cm_U\simeq  H^0(\cp,\cb)/H^0(\cp,\cs),
\eqno(19)
$$
that follows from the definition (5) of the moduli space $\cm_U$
and the twistor correspondence briefly described in \S\,2.3. The
description of $\cm_U$ in terms of $\cg$-valued (0,1)-forms
$\bar B$ on $\cp$ is called the {\it Dolbeault description} of $\cm_U$.

Now let us consider the set $\ker\ff = \ff^{-1}(e)$, $e\in H^1(\cp,\cs)$.
It consists of such elements from $H^1(\cp,\ch)$ that are mapped into the
class $e\in H^1(\cp,\cs)$ of smoothly trivial complex vector bundles over
$\cp$ that are holomorphically trivial on any projective line
$\C P^1_x\hra\cp, x\in U$.  Therefore, the set  $\ker\ff $ is the
moduli space of holomorphic
vector bundles $\tilde \ce$ that are diffeomorphic  to the bundle
$\tilde\ce_0$ from the class $e\in H^1(\cp,\cs)$. For any representative
$\cf=\{\cf_{12}, \cf_{12}^{-1}\}\in Z^1(\fu,\ch)\subset Z^1(\fu,\cs)$
of the set $\ker\ff $ one can find a decomposition
$$
\cf_{12}=\psi_1^{-1}(x,\l ) \psi_2 (x,\l^{-1} ),
\eqno(20)
$$
where $\psi_1$, $\psi_2$ are smooth $G$-valued functions on $\U_1, \U_2$
that are holomorphic on ${\C}P^1_x\hra\cp, x\in U$. Note that
$\psi =\{\psi_1, \psi_2\}\in C^0(\fu,\cs)$.

It follows from the exact sequence (18) that
$$
\ker\ff\simeq H^0(\cp,\cb)/H^0(\cp,\cs).
\eqno(21)
$$
Therefore we have the bijection
$$
\cm_U\simeq\ker\ff ,
\eqno(22)
$$
and the description of $\cm_U$ in terms of transition matrices
$\cf\in\ker\ff$ is called the {\it \v{C}ech description} of the
moduli space $\cm_U$.

Let us collect the bijections (19), (21) and (22) in the following
table:

\bigskip
\begin{tabular}{lcr}
{\sf the Dolbeault description}&
{\sf the moduli space }&
{\sf the \v{C}ech description}\\
& {\sf of self-dual gauge fields}& \\
$H^{0,1}_{\bar\p_{\hat B}}(\cp)\supset H^0(\cp,\cb)/H^0(\cp,\cs)$&
$\simeq\quad\cm_U\quad\simeq$&$\ker\ff \subset H^1(\cp,\ch)$,\\
\end{tabular}

\bigskip

\noindent
where  $H^{0,1}_{\bar\p_{\hat B}}(\cp)$ is a Dolbeault 1-cohomology set
defined as a set of orbits of the group $H^0(\cp,\fs)$ in the set
$H^0(\cp,\fb)$ and $\fb$ is the sheaf of $\cg$-valued $(0,1)$-forms
$\hat B$ on $\cp$ such that $\bar\p^2_{\hat B}=0$.

\ewmsection{Infinitesimal symmetries of the SDYM equations}

{\em We can now use the results of the previous sections to study
symmetries of the SDYM equations.  Cohomological description of the
moduli space of self-dual gauge fields simplifies the problem of
finding symmetries of the SDYM equations and clarifies the geometric
meaning of these symmetries.  Namely, in the \v{C}ech approach, to
solutions of the SDYM equations there correspond holomorphic
$G$-valued functions  $\cf_{12}$ (1-cocycles) on the overlap
$\U_{12}$ of the open sets $\U_1$, $\U_2$ covering the twistor space
$\cp$.  Therefore any holomorphic perturbation of $\cf_{12}$
determines a tangent vector on the solution space of the SDYM
equations.  In \S\,4.2 we define these infinitesimal holomorphic
transformations of $\cf_{12}$ by multiplying $\cf_{12}$ on holomorphic
$\cg$-valued matrices $\th_{12}$, $\th_{21}$ defined on $\U_{12}$.
Then, using a solution of the infinitesimal variant of the
Riemann-Hilbert problem from \S\,4.3,  we proceed in \S\,4.4 to the
Dolbeault description and define a transformation of the flat
(0,1)-connection $\bar B$.  Finally we introduce the algebra
$C^1(\fu ,\dot\ch )$ of 1-cochains of $\cp$ with values in the
sheaf $\dot\ch$ of $\cg$-valued holomorphic functions on $\cp$ and,
using the Penrose-Ward correspondence, we describe in \S\,4.5 the
action of the algebra $C^1(\fu ,\dot\ch )$ on self-dual gauge potentials.}

\ewmsubsection{Action of the group $C^1(\fu, \ch )$ on the space
               $Z^1(\fu, \ch )$}

The group $C^1(\fu, \ch )$ and the space $Z^1(\fu, \ch )$ have been
described in \S\,3.1. Let us define the action $\rho$ of
$C^1(\fu, \ch )$ on $Z^1(\fu, \ch )$ by the formula
$$
(\rho_hf)_{12}=h_{12}f_{12}h_{21}^{-1},
\eqno(23)
$$
where $h=\{h_{12}, h_{21}\}\in C^1(\fu, \ch )$ , $f=\{f_{12}, f_{12}^{-1}\}
\in Z^1(\fu, \ch )$. It is clear that for an arbitrary cocycle
$f=\{f_{12}, f_{21}\}\in Z^1(\fu, \ch )$, one can always find a cochain
$\{h_{12}, h_{21}\}\in C^1(\fu, \ch )$ such that $f_{12}=h_{12} h_{21}^{-1}$,
$f_{21}=h_{21} h_{12}^{-1}$, i.e., the group $C^1(\fu, \ch )$  acts
transitively on $Z^1(\fu, \ch )$. The stability subgroup of the trivial
cocycle $f^0=1$ is
$$
C_{\triangle}(\fu, \ch )=\{\{h_{12}, h_{21}\}\in C^1(\fu, \ch ):
h_{12}=h_{21}\}.
$$
Therefore,  $Z^1(\fu, \ch )$ is a homogeneous space,
$$
Z^1(\fu, \ch )=C^1(\fu, \ch )/ C_{\triangle}(\fu, \ch ).
$$

\ewmsubsection{Action of the algebra $C^1(\fu, \dot\ch )$ on the space
               $Z^1(\fu, \ch )$}

Let us denote by $\dot\ch$ the sheaf of holomorphic sections of the
trivial bundle $\cp\times\cg$, where $\cg$ is the Lie algebra of a
Lie group $G$. Denote by $\dot\cs$ the sheaf of smooth partially
holomorphic sections of the bundle $\cp\times\cg$, i.e., such smooth
maps $\phi : \cp\to\cg$ that $\p_{\bar\l}\phi =0$ in the local coordinates
$\{x^{\mu},\l ,\bar\l\}$ on $\cp$.

We consider the infinitesimal form of the action (23). Substituting
$h_{12}=\exp (\th_{12})\simeq 1+\th_{12}$, $h_{21}=\exp (\th_{21})
\simeq 1+\th_{21}$, we have
$$
\d_{\th}\cf_{12}=\th_{12}\cf_{12} - \cf_{12}\th_{21},
\eqno(24)
$$
where $\th =\{\th_{12}, \th_{21}\}\in C^1(\fu, \dot\ch )$, $\cf =\{\cf_{12},
\cf_{12}^{-1}\}\in Z^1(\fu, \ch )$. Here and in what follows as
$\cf =\{\cf_{12}, \cf_{12}^{-1}\} $ we take representatives of the space
$\ker\ff$ (see \S\,3.3), i.e., such cocycles $\cf_{12}$ that admits the
decomposition  (20).

\ewmsubsection{The map $\phi : C^1(\fu, \dot\ch )\to C^0(\fu, \dot\cs )$}

Now we construct the following $\cg$-valued function:
$$
\Phi_{12}(\th )=\psi_1(\d_{\th}\cf_{12})\psi_2^{-1},
\eqno(25)
$$
where $\{\psi_1, \psi_2\}\in C^0(\fu,\cs)$ and $\cf_{12}=
\psi_1^{-1}\psi_2$. Then one can check that
$$
\Phi_{21}=-\Phi_{12}
$$
and $\Phi_{12}$ is a smooth $\cg$-valued function on $\U_{12}$
such that $\p_{\bar\l}\Phi_{12}=0$ in the local coordinates
$\{x^\mu ,\l , \bar\l\}$ on $\U_{12}$. Therefore, $\Phi =\{\Phi_{12},
\Phi_{21}\}\in Z^1(\fu,\dot\cs)$.

It can be shown that $H^1(\cp ,\dot\cs )=0$, since $\dot\cs$ is the sheaf
of smooth $\cg$-valued functions on $\cp$ that are holomorphic on
$\C P^1\hra\cp$. Therefore, each 1-cocycle with values in $\dot\cs$
is a 1-coboundary, and we have
$$
\Phi_{12}(\th ) = \phi_1(\th ) - \phi_2(\th ),
\eqno(26)
$$
where $\phi (\th )=\{\phi_1(\th ), \phi_2(\th )\}\in C^0(\fu,
\dot\cs )$.

Notice that the splitting (26) defined for any $\th\in C^1(\fu ,
\dot\ch )$ is not unique. Namely, as a 0-cochain from 
$C^0(\fu,\dot\cs )$ instead of $\phi (\th )$ one can also take
$$
\tilde\phi (\th )=\{\phi_1(\th )+\vp_1, \phi_2(\th )+\vp_2\},
$$
where $\vp_1=\vp_2$ on $\U_{12}$, i.e., $\vp =\{\vp_1,\vp_2\}\in
H^0(\cp,\dot\cs)$. Let us fix $\vp\in H^0(\cp,\dot\cs)$ for each
$\th\in C^1(\fu , \dot\ch )$, then the splitting (26) defines
a subspace $\phi (C^1(\fu , \dot\ch ))$ in $C^0(\fu,\dot\cs )$.
It can be checked that
$$
\phi ([\th, \tilde\th])= [\phi (\th ), \phi (\tilde\th )]=
\{[\phi_1(\th ), \phi_1(\tilde\th )], [\phi_2(\th ), 
\phi_2(\tilde\th)]\} \in C^0(\fu,\dot\cs )
$$
{}for any $\th, \tilde\th\in C^1(\fu , \dot\ch )$. Therefore, the map
$\phi : C^1(\fu, \dot\ch )\to C^0(\fu, \dot\cs )$ is a homomorphism.

\ewmsubsection{Action of the algebra $C^1(\fu, \dot\ch )$ on the space
               $H^0(\cp, \cb )$}

Using the action (24) and the homomorphism $\phi$, we obtain an action
$$
\d_\th\psi_1=-\phi_1(\th )\psi_1,\quad
\d_\th\psi_2=-\phi_2(\th )\psi_2,
\eqno(27)
$$
of the algebra $C^1(\fu , \dot\ch )$ on a 0-cochain $\{\psi_1,\psi_2\}
\in
C^0(\fu, \cs)$ such that $\cf_{12}=\psi_1^{-1}\psi_2$.

By definition, for $\bar B=\{\bar B^{(1)}, \bar B^{(2)}\}\in H^0(\cp,
\cb)$
we have
$$
\bar B^{(1)} = - (\bar\p\psi_1)\psi_1^{-1}\ \mbox{on}\ \U_1, \quad
\bar B^{(2)} = - (\bar\p\psi_2)\psi_2^{-1}\ \mbox{on}\ \U_2, \quad
\bar B^{(1)} = \bar B^{(2)}\ \mbox{on}\ \U_{12}=\U_1\cap\U_2.
$$
Therefore, the action of $C^1(\fu ,\dot\ch)$ on $H^0(\cp,\cb)$ has
the form
$$
\d_\th\bar B^{(1)} = \bar\p\phi_1(\th ) + [\bar B^{(1)}, 
\phi_1(\th )],
\eqno(28a)
$$
$$
\d_\th\bar B^{(2)} = \bar\p\phi_2(\th ) + [\bar B^{(2)}, 
\phi_2(\th )].
\eqno(28b)
$$
The transformations (28) look like infinitesimal gauge 
transformations
$$
\d_\vp\bar B=\bar\p\vp + [\bar B,\vp ],
\eqno(29)
$$
where $\vp$ is an element of the Lie algebra
$H^0(\cp,\dot\cs )\simeq\fg_U$ of the gauge group
$H^0(\cp,\cs )\simeq\frg_U$. But for $\phi (\th )=\{\phi_1(\th ),
\phi_2(\th )\}\in C^0(\fu,\dot\cs )$ we have  $\phi_1(\th )\ne
\phi_2(\th )$
on $\U_{12}$, and the transformations (28) differ from (29).

\ewmsubsection{Action of the algebra $C^1(\fu, \dot\ch )$ on the space
               $\ca_U$}

Recall that we consider a self-dual 4-manifold $M$, the twistor space 
$\cz$ of which is a complex 3-manifold, and the SDYM equations (1) on 
$M$. To describe infinitesimal symmetries of the SDYM equations, we 
take an open ball $U\subset M$ and the twistor space $\cp$ of $U$ 
that is covered by two coordinate patches $\U_1$ and $\U_2$ (see 
\S\,2.2).

The twistor correspondence gives us the following relation between a
self-dual connection $A=A_\mu dx^\mu$ on the complex vector bundle
$\ce$ over $U$ and a flat $(0,1)$-connection $\bar B=\{\bar B^{(1)},
B^{(2)}\}$ on the bundle $\tilde\ce_0=\pi^*\ce$:
$$
\bar B^{(1)}_1=A_{\bar y}-\l A_z, \
\bar B^{(1)}_2=A_{\bar z}+\l A_y, \ \bar B^{(1)}_3=0\ \mbox{on}\ 
\U_1,
\eqno(30a)
$$
$$
\bar B^{(2)}_1=\z A_{\bar y}- A_z, \
\bar B^{(2)}_2=\z A_{\bar z}+ A_y, \ \bar B^{(2)}_3=0\ \mbox{on}\ 
\U_2,
\eqno(30b)
$$
where $y=x^1+ix^2,\ z=x^3-ix^4, \bar y=x^1-ix^2,\ \bar z=x^3+ix^4$
are complex coordinates on $U$.

One can always choose such local frames $\{\bar V^{(1)}_a\},\
\{\bar V^{(2)}_a\}$ of the bundle $T^{0,1}$ over $\U_1$, $\U_2$,
respectively, that $[\bar V^{(1)}_a, \bar V^{(1)}_b]=0,\
[\bar V^{(2)}_a, \bar V^{(2)}_b]=0,\ \bar V^{(1)}_3=\p_{\bar\l},\
\bar V^{(2)}_3 =\p_{\bar\z}$ and on the intersection
$\U_{12}=\U_1\cap\U_2$ the local frames are connected by
the formulae~\cite{Pen, AHS, W80}
$$
\bar V^{(1)}_1=\l\bar V^{(2)}_1,\quad
\bar V^{(1)}_2=\l\bar V^{(2)}_2,\quad
\bar V^{(1)}_3=-\bar\l^2\bar V^{(2)}_3.
$$

{}From (27), (28) we obtain the following action of the algebra
$C^1(\fu,\dot\ch)$ on the space $\ca_U$ of solutions to the SDYM
equations on $U$:
$$
\d_\th A_y = \oint_{S^1}\frac{d\l}{2\pi i\l}(\bar V^{(2)}_2+\bar
B^{(2)}_2)\phi_2(\th ),\quad
\d_\th A_z = -\oint_{S^1}\frac{d\l}{2\pi i\l}(\bar V^{(2)}_1+
\bar B^{(2)}_1)\phi_2(\th ),
$$
$$
\d_\th A_{\bar y} = \oint_{S^1}\frac{d\l}{2\pi i\l}(\bar V^{(1)}_1+
\bar B^{(1)}_1)\phi_1(\th ),\quad
\d_\th A_{\bar z} = \oint_{S^1}\frac{d\l}{2\pi i\l}(\bar V^{(1)}_2+
\bar B^{(1)}_2)\phi_1(\th ),
\eqno(31)
$$
where $S^1=\{\l\in\C P^1: |\l |=1\}$.

\ewmsection{Conclusion}

The space of local solutions to the SDYM equations on a self-dual
4-manifold $M$ has been considered.  Choosing the concrete self-dual 
4-manifold (e.g. $S^4$, $T^4$,\ ...) or imposing some boundary 
conditions on gauge fields, one can obtain instantons, monopoles or 
other special solutions of the SDYM equations, the moduli spaces of 
which are discussed in the talk by S.T.Tsou~\cite{tsou}. Our purpose 
was to describe the moduli space and symmetries of {\em local} 
solutions to the SDYM equations.  The use of twistor correspondence 
and cohomologies reveals the geometric meaning of symmetries of the 
SDYM equations, which may help in quantizing the SDYM model.

\section*{Acknowledgements}

The author thanks the conference organizers Frances Kirwan, Sylvie 
Paycha and Sheung Tsun Tsou for their invitation, kind hospitality in 
Oxford and for creating a very pleasant and stimulating atmosphere.  
The author is grateful to Sylvie Paycha for reading the manuscript 
and valuable remarks.

\end{document}